\documentclass{new_tlp}

\usepackage{url,graphicx}

\newcommand\bcmdtab{\noindent\bgroup\tabcolsep=0pt%
  \begin{tabular}{@{}p{10pc}@{}p{20pc}@{}}}
\newcommand\ecmdtab{\end{tabular}\egroup}

  \title[Combinatorial Search With Picat]
        {Combinatorial Search With Picat}

  \author[N.-F. Zhou]
         {NENG-FA ZHOU\\
         Brooklyn College and Graduate Center \\
         The City University of New York\\
         \email{nzhou@acm.org}}

\jdate{February 2014}
\pubyear{2014}
\pagerange{\pageref{firstpage}--\pageref{lastpage}}
\doi{XXX}

\begin{document}

\label{firstpage}

\maketitle

  \begin{abstract}
Picat, a new member of the logic programming family, follows a different doctrine than Prolog in offering the core logic programming concepts: arrays and maps as built-in data types; implicit pattern matching with explicit unification and explicit non-determinism; functions for deterministic computations; and loops for convenient scripting and modeling purposes. Picat provides facilities for solving combinatorial search problems, including a common interface with CP, SAT, and MIP solvers, tabling for dynamic programming, and a module for planning. Picat's planner module, which is implemented by the use of tabling, has produced surprising and encouraging results. Thanks to term-sharing and resource-bounded tabled search, Picat overwhelmingly outperforms the cutting-edge ASP and PDDL planners on the planning benchmarks used in recent ASP competitions.
  \end{abstract}



\section{Introduction}
Picat is a simple, and yet powerful, logic-based multi-paradigm programming language. The desire for a logic-based general-purpose programming language that is as powerful as Python for scripting, and on a par with OPL \cite{Hentenryck02} and MiniZinc \cite{NethercoteSBBDT07} for modeling combinatorial problems, led to the design of Picat. Early attempts to introduce arrays and loops into Prolog for modeling failed to produce a satisfactory language: most noticeably, array accesses are treated as functions only in certain contexts; and loops require the declaration of global variables in ECLiPSe \cite{Schimpf02} and local variables in B-Prolog \cite{Zhou12}. 

Picat departs from Prolog in many aspects, including the successful introduction of arrays and loops. Picat uses pattern-matching rather than unification in the selection of rules. Unification might be a natural choice in Horn clause resolution \cite{kowalski1971} for theorem proving, but its power is rarely needed for general programming tasks. Pattern-matching rules are fully indexed, and therefore Picat can be more scalable than Prolog. Unification can be considered as an equation over terms \cite{Colmerauer84}, and just like constraints over finite domains, Picat supports unification as an explicit call.

Non-determinism, a powerful feature of logic programming, makes concise solutions possible for many problems, including simulation of non-deterministic automata, parsers of ambiguous grammars, and search problems. Nevertheless, non-determinism is not needed for deterministic computations. In Prolog, Horn clauses are backtrackable by default. As it is undecidable to detect determinism in general \cite{Debray88}, programmers tend to excessively use the cut operator to prune unnecessary clauses. Picat supports explicit non-determinism, which renders the cut operator unnecessary. Rules are deterministic unless they are explicitly denoted as backtrackable. 

Picat supports functions, like many other logic-based languages, such as Curry \cite{Hanus13}, Erlang \cite{Armstrong13}, and Mozart-Oz \cite{RoyH2004}. In Prolog, it's often that queries fail, but the system gives no clue about the source of the failure. Functions should be used instead of relations, unless multiple answers are required. Functions are more convenient to use than predicates because (1) functions are guaranteed to succeed with a return value; (2) function calls can be nested; and (3) the directionality of functions enhances the readability.

Many combinatorial problems can be formulated as constraint satisfaction problems (CSPs). There are three kinds of systematic solvers for solving CSPs, namely, Constraint Programming (CP), Mixed Integer Programming (MIP), and SAT solving. CP uses constraint propagation to prune search spaces, and uses heuristics to guide search \cite{Rossi06}. MIP relies on LP relaxation and branch-and-cut to find optimal integer solutions \cite{Appa10}. SAT performs unit propagation and clause learning to prune search spaces, and employs heuristics and learned clauses to perform non-chronological backtracking \cite{MalikZ09}. No solver is superior all the time; sometimes, extensive experimentation is necessary to find a suitable solver.

Picat provides a common interface with CP, SAT, and MIP solvers for solving CSPs. For each solver, Picat provides a separate module of built-ins for creating decision variables, specifying constraints, and invoking the solver. The common interface allows for seamless switching from one solver to another. The basic language constructs, such as arrays and loops, make Picat a powerful modeling language for these solvers.

Tabling \cite{warren92} can be used to cache the results of certain calculations in memory and reuse them in subsequent calculations through a quick table lookup. As computer memory grows, tabling is becoming increasingly important for offering dynamic programming solutions for many problems. Picat's tabling system is inherited from B-Prolog \cite{zhou08tab}. 

Picat has a planner module. For a planning problem, the programmer only needs to specify conditions on the final states and the set of actions, and to call the planner on an initial state to find a plan or an optimal plan. The planner, which is implemented by the use of tabling, performs a state-space search and tables every state that is encountered during search. 

A joint effort by the system and the programmer is needed to deal with the state explosion problem. The Picat system stores all structured ground terms in a table, so ground terms that are shared by states are only tabled once. The enhanced {\it hash-consing} technique \cite{ZhouH12} also stores hash codes in order to speed up computation of hash codes and equality tests of terms. The Picat system also performs {\it resource-bounded tabled search}, which prunes parts of the search space that cannot lead to acceptable plans. In order to exploit these techniques, the programmer needs to design a good representation for states that facilitates sharing and removes symmetries. For certain problems, the programmer can also employ domain knowledge and heuristics to help prune the search space. 

Picat's planner has produced surprising and encouraging results. It overwhelmingly outperforms the cutting-edge ASP and PDDL planners on many benchmarks used in recent ASP and IPC competitions. The Picat encodings of the benchmarks, which are as compact as the ASP and PDDL encodings, are available at \url{picat-lang.org}.

This paper gives an overview of Picat's facilities for combinatorial search. It also offers a glimpse of the language features. The readers are referred to \cite{PicatGuide,Kjellerstrand14} for the details of the language.

\section{\label{sec:overview}An Overview of Picat}
Picat follows a different doctrine than Prolog in offering the core logic programming concepts. This section gives a brief overview of Picat's basic language constructs. The facilities for combinatorial search, including tabling, solver modules for CSPs, and a module for planning, will be described later. Other features of Picat, which are not covered in this overview, include assignments, global maps, action rules for defining event-driven actors, a simple module system, modules for everyday programming tasks ({\tt basic}, {\tt math}, {\tt io}, {\tt util}, and {\tt os}), and a module for probabilistic reasoning and learning with PRISM \cite{prism:website}. 

\subsection{Logic Variables and Data Types}
A logic variable is a value holder, and a value is a term, which can be another variable. In addition to the basic data types in Prolog, Picat also provides strings, arrays, and maps. A double-quoted string is represented as list of single-character atoms, and all of the built-ins on lists, such as the concatenation function \verb-++-, can also be applied to strings. An \emph{array} takes the form \texttt{\{$t_1$,$\ldots$,$t_{n}$\}}. In the current implementation, an array is a special structure with the name \texttt{`\{\}'}. A \emph{map} is a hash-table that is represented as a structure, containing a set of key-value pairs. 

Picat allows function calls in arguments. For this reason, it requires structures to be preceded with a dollar sign in order for them to be treated as data. For example, \verb+$student(mary,cs,3.8)+ is a structure, not a function call. Special structures, such as {\tt (A,B)} and {\tt (A;B)}, as well as head patterns, are not required to have a dollar sign.

For each type, Picat provides a set of built-in functions and predicates. The index notation \texttt{$X$[$I$]}, where $X$ references a compound value and $I$ is an integer expression, is a special function that returns the component of $X$ at index $I$. The index of the first element of a list or a structure is 1. 

Picat also allows OOP notations for accessing attributes and for calling predicates and functions. The notation \texttt{$A_1.f(A_2,\ldots,A_k)$} is the same as \texttt{$f(A_1,A_2,\ldots,A_k)$}, unless $A_1$ is an atom, in which case $A_1$ must be a module qualifier for $f$.  The notation $A.Attr$, where $Attr$ is not in the form $f(\ldots)$, is the same as the function call \texttt{get$(A,Attr)$}. A structure is assumed to have two attributes called \texttt{name} and \texttt{length}. 

\subsection{Pattern-matching Rules and Explicit Non-determinism}
In Picat, predicates and functions are defined with pattern-matching rules. Picat has two types of rules: the non-backtrackable rule $Head, Cond\ $\verb+=>+$\ Body$, and the backtrackable rule $Head, Cond\ $\verb+?=>+$\ Body$. In a predicate definition, the $Head$ takes the form $p(t_1,\ldots,t_n)$, where $p$ is called the predicate name, and $n$ is called the arity. The condition $Cond$, which is an optional goal, specifies a condition under which the rule is applicable. For a call $C$, if $C$ matches $Head$ and $Cond$ succeeds, then the rule is said to be \emph{applicable} to $C$. For a head in which a variable occurs more than once, such as {\tt p(X,X)}, a call matches the pattern only if the arguments are identical. Unlike the pattern matching that is used in concurrent logic languages \cite{SHA89}, a call fails rather than freezes when it contains insufficiently instantiated arguments.

A pattern can contain \emph{as-patterns} in the form \texttt{$V$@$Pattern$}, where $V$ is a new variable in the head, and $Pattern$ is a non-variable term. The as-pattern \texttt{$V$@$Pattern$} is the same as \texttt{$Pattern$} in pattern matching, but after pattern matching succeeds, $V$ is made to reference the term that matches $Pattern$. As-patterns can be used to avoid re-constructing existing terms.

When applying a rule to call $C$, Picat rewrites $C$ into $Body$. If the used rule is non-backtrackable, then the rewriting is a commitment, and the program can never backtrack to $C$. However, if the used rule is backtrackable, then the program will backtrack to $C$ once $Body$ fails, meaning that $Body$ will be rewritten back to $C$, and the next applicable rule will be tried on $C$. 

Pattern matching does not change the status of the variables in a call. In order to bind a variable $X$ in a call to a value $Y$, users can call the unification $X=Y$. While it is not illegal to bind variables in $Cond$, $Cond$ normally contains only tests, and all unification calls should be written in $Body$. For example,
\begin{verbatim}
    member(X,[Y|_]) ?=> X=Y.
    member(X,[_|L]) => member(X,L).
\end{verbatim}
The first rule is backtrackable. This predicate can be used to retrieve elements from a given list one by one through backtracking.

\subsection{Functions}
A function call always succeeds with a return value, unless an exception occurs. Functions are defined with non-backtrackable rules in the form $F$\verb+=+$Exp, Cond\ $\verb+=>+$\ Body$, where $F$ is a function pattern in the form $f(t_1,\ldots, t_n)$, and $Exp$ is an expression. When both $Cond$ and $Body$ are {\tt true}, the rule can be written as $F$\verb+=+$Exp$. Functions are compiled into predicates.

A function call never fails due to failures in pattern matching. If no rule is applicable to a function call, then the system throws an {\tt unresolved\_function\_call} exception.

\subsection{Loops and List Comprehension}
Picat allows loops in rule bodies.  Loops are compiled into tail-recursive predicates. A \texttt{foreach} statement takes the form
\begin{tabbing}
aa \= aaa \= aaa \= aaa \= aaa \= aaa \= aaa \kill
\> \texttt{foreach ($E_1$ in $D_1$, $Cond_1$, $\ldots$, $E_n$ in $D_n$, $Cond_n$)}  \\
\> \> $Goal$ \\
\>  \texttt{end} 
\end{tabbing}
where each iterator, $E_i\ in\ D_i$, can be followed by an optional condition $Cond_i$.  Within each iterator, $E_i$ is an iterating pattern, and $D_i$ is an expression that gives a compound value. The \texttt{foreach} statement means that $Goal$ is executed for every possible combination of values $E_1 \in D_1$, $\ldots$, $E_n \in D_n$ that satisfies the conditions \texttt{$Cond_1$}, $\ldots$, \texttt{$Cond_n$}. 

Picat adopts the following simple scoping rule: {\it variables that occur only in a loop, but do not occur before the loop in the outer scope, are local to each iteration of the loop}. For example,
\begin{verbatim}
    p(A) =>                                    
        q(X),
        foreach (I in 1 .. A.length)            
            A[I] = (X,Y)                      
        end.                                   
\end{verbatim}
The loop unifies each element {\tt A[I]} of array {\tt A} with a tuple {\tt (X,Y)}, where {\tt X} is global and is the same for every iteration, and {\tt Y} is local and is new to each iteration.

A list comprehension, which takes the following form, is a special functional notation for creating lists:
\begin{tabbing}
aa \= aaa \= aaa \= aaa \= aaa \= aaa \= aaa \kill
\> \> \texttt{[$T$ : $E_1$ \texttt{in} $D_1$, $Cond_1$, $\ldots$, $E_n$ in $D_n$, $Cond_n$]} 
\end{tabbing}
where $T$ is an expression. This list comprehension means that for every tuple of values $E_1 \in D_1$, $\ldots$, $E_n \in D_n$, if the conditions are true, then the value of $T$ is added into the list.

Picat supports the assignment operator {\tt :=}. The assignment $X$ {\tt :=} $Y$, where $X$ is a variable, does not actually assign the value of $Y$ to $X$. It creates a new variable for $X$ to hold the value of $Y$. After the assignment, whenever $X$ is accessed in the body, the new variable is accessed. With assignments, a list comprehension can be easily compiled into a {\tt foreach} loop that uses an assignment to accumulate the list.

Loops are convenient for scripting and modeling purposes. Figure \ref{fig:loops} gives three example functions that would be difficult to write without using loops or list comprehension.

\begin{figure}[tb]
\begin{center}
\begin{verbatim}
        power_set([]) = [[]].
        power_set([H|T]) = P1++P2 =>
            P1 = power_set(T),
            P2 = [[H|S] : S in P1].

        perm([]) = [[]].
        perm(Lst) = [[E|P] : E in Lst, P in perm(Lst.delete(E))].

        matrix_multi(A,B) = C =>
            C = new_array(A.length,B[1].length),
            foreach (I in 1..A.length, J in 1..B[1].length)
                C[I,J] = sum([A[I,K]*B[K,J] : K in 1..A[1].length])
            end.
\end{verbatim}
\end{center}
\caption{\label{fig:loops}Three example functions in Picat.}
\end{figure}

\section{A Common Interface With CP, SAT, and MIP}
Picat provides three solver modules, including \texttt{cp}, \texttt{sat} and \texttt{mip}. Each of the three solver types has its strengths and weaknesses. In reality, extensive experimentation is required in order to determine a proper model and to find a suitable solver. All of the three modules implement the same interface, which makes it seamless to switch from one solver to another.

\subsection{The Common Interface}
The common interface consists of built-ins for creating decision variables, specifying constraints, and invoking the imported solver. In order to use a solver, users must first import the module.

A decision variable is a logic variable with a domain. The domain constraint {\tt $Vs$ :: $D$} narrows the domains of the variables $Vs$ to $D$. $Vs$ is a variable, a list of variables, or an array of variables. $D$ is an expression that gives a list of integers. 

An arithmetic constraint takes the form of $E_1\ R\ E_2$, where $E_1$ and $E_2$ are two arithmetic expressions, and $R$ is one of the following constraint operators: \verb+#=+ (equal), \verb+#!=+ (not equal), \verb+#>=+, \verb+#>+, \verb+#=<+ (\verb+#<=+), and \verb+#<+. An arithmetic expression is made from integers, domain variables, and built-in arithmetic functions.

A basic Boolean expression is made from constants (0 and 1), Boolean variables, and the following operators: \verb+#/\+ (and), \verb+#\/+ (or), \verb+#~+ (not), \verb+#^+ (xor), \verb+#<=>+ (equivalent), and \verb+#=>+ (implication).  An extended Boolean expression can also include arithmetic and domain constraints as operands. In particular, the constraint \verb+B #<=> (E1 #= E2)+ is called a {\it reification} constraint, which uses a Boolean variable {\tt B} to indicate the satisfiability of the arithmetic constraint \verb+E1 #= E2+.

A \emph{table constraint}, or an \emph{extensional constraint}\index{extensional constraint}, over a tuple of variables specifies a set of tuples that are allowed ({\tt table\_in}) or disallowed ({\tt table\_notin}) for the variables.

The interface also contains the commonly used global constraints, such as the {\tt all\_different}, {\tt element}, {\tt circuit}, and {\tt cumulative} constraints.

The built-in predicate \texttt{solve($Options$, $Vars$)} calls the imported solver to label the variables $Vars$ with values, where $Options$ is a list of options for the solver. When the option {\tt min($E$)} or {\tt max($E$)} is included, the solver returns an optimal answer.

Figure \ref{fig:queens} gives a program for the N-queens problem.
\begin{figure}[hbt]
\begin{center}
\begin{verbatim}
        import cp.

        queens(N, Q) =>
            Q = new_list(N),
            Q :: 1..N,
            all_different(Q),
            all_different([$Q[I]-I : I in 1..N]),
            all_different([$Q[I]+I : I in 1..N]),
            solve([ff],Q).
\end{verbatim}
\end{center}
\caption{\label{fig:queens}A Picat program for N-queens.}
\end{figure}
\vspace*{-3mm}

\subsection{Implementation of the Solver Modules}
An underlying solver is used for each of the solver modules: the {\tt cp} module uses a solver inherited from B-Prolog; the {\tt sat} module uses Lingeling\footnote{fmv.jku.at/lingeling} on Linux and MiniSat\footnote{minisat.se/} on Windows; the {\tt mip} module uses GLPK\footnote{www.gnu.org/software/glpk/}.

For the {\tt cp} module, constraints are compiled into propagators that are defined in the AR (Action Rules) language \cite{zhou06ar}, which are compiled further into abstract machine instructions. The abstract machine provides native support for fast propagation. In particular, it stores propagators on the stack for fast context switching and provides specialized instructions for encoding commonly used propagators \cite{zhou06ar}. The solver, which has competed in numerous solver competitions since 2005, is robust and efficient. For example, Picat solves the N-queens problem for N=1500 in less than 10 seconds on an Intel i5 machine.

For the {\tt sat} module, constraints are compiled into a logic formula in the conjunctive normal form (CNF) for the underlying SAT solver. Picat employs the so called {\it log-encoding} for compiling domain variables and constraints. For a domain variable,  $\lceil log_2(n)\rceil$ Boolean variables are used, where $n$ is the maximum absolute value of the domain. If the domain contains both negative and positive values, then another Boolean variable is used to encode the sign. Each combination of values of these Boolean variables represents a valuation for the domain variable. If there are holes in the domain, then disequality ($\neq$) constraints are generated in order to disallow assignments of those hole values to the variable. Equality and disequality constraints are flattened to two types of primitive constraints in the form of $x>y$ and $x+y=z$, which are compiled further into logic comparators and adders in CNF. For other types of constraints, clauses are generated in order to disallow conflict values for the variables. 

The same log-encoding is used by the FlatZinc SAT compiler \cite{Huang08}. Log-encoding has less propagation power than {\it direct} and {\it support} encodings for certain constraints \cite{Gavanelli07}, but is much more compact than other encodings, including the {\it order} encoding which is adopted by the Sugar \cite{TamuraTKB09} and the BEE \cite{MetodiC12} compilers. The {\tt sat} module has solved many problems that are hard to solve with the {\tt cp} module.

The MIP solver is still the first choice for many Operations Research applications \cite{Appa10}. For the {\tt mip} module, constraints are compiled into inequality ($\le$) constraints. The compilation follows the standard textbook recipe. For example, the constraint \verb+X #!= Y+ is first translated to \verb&X #=< Y-1 #\/ X #>= Y+1&, which is then translated to \verb&B1 #\/ B2&, where 
\begin{verbatim}
    B1 #<=> (X #=< Y-1)
    B2 #<=> (X #>= Y+1)
\end{verbatim}
The reification constraint \verb&B #<=> (X #=< Y)& is compiled to \verb&X-Y-M1*(1-B) #=< 0& and \verb&Y-X+1-M2*B #=< 0&, where {\tt M1} and {\tt M2} are constants:
\begin{verbatim}
    M1 = ubd(X)-lbd(Y)+1
    M2 = ubd(Y)-lbd(X)+2
\end{verbatim}
where {\tt lbd(X)} is the lower bound of the domain of {\tt X}, and {\tt ubd(X)} is the upper bound.

\section{Tabling for Dynamic Programming}
The idea of tabling is to store tabled calls and their answers in a table, and to use the answers to resolve subsequent variant calls. This idea has been used in functional and logic programming for several decades, dating back to \cite{Michie68} and \cite{Tamaki86}. As computer memory grows and advanced implementation techniques are invented, tabling is becoming increasingly important for offering dynamic programming solutions for many problems. 

Picat's tabling system is inherited from B-Prolog. In order to have all of the calls and answers of a predicate or a function tabled, users just need to add the keyword \texttt{table} before the first rule. Picat supports mode-directed tabling for dynamic programming problems \cite{GuoG08}.  For a tabled predicate, users can give a \emph{table mode declaration} in the form {\tt table($M_{1},M_{2},\ldots,M_{n}$)}, where each $M_{i}$ is one of the following: a plus-sign (+) indicates input, a minus-sign (-) indicates output, \texttt{max} indicates that the corresponding argument is maximized, and \texttt{min} indicates that the corresponding argument is minimized. The last mode, $M_{n}$, can be \texttt{nt}, which indicates that the argument is not tabled. Two types of data can be passed to a tabled predicate as an \texttt{nt} argument: (1) global data that are the same to all of the calls of the predicate, and (2) data that are functionally dependent on the input arguments. 

When a table mode declaration is provided, Picat only tables the current best answer for each tuple of input arguments. Picat uses linear tabling \cite{zhou08tab} to iteratively evaluate looping calls until an optimal answer is found. Mode-directed tabling assumes that the objective function grows or declines monotonically.

For example, the following tabled predicate searches for a path with the maximum total sum from top to bottom in a triangle. 
\begin{verbatim}
    table (+,+,max,nt) 
    path(Row,Col,Sum,Tri),Row==Tri.length => Sum=Tri[Row,Col].
    path(Row,Col,Sum,Tri) ?=> 
        path(Row+1,Col,Sum1,Tri),
        Sum = Sum1+Tri[Row,Col].
    path(Row,Col,Sum,Tri) => 
        path(Row+1,Col+1,Sum1,Tri),
        Sum = Sum1+Tri[Row,Col].    
\end{verbatim}
The triangle, which is represented as an array of arrays, is passed as an {\tt nt} argument. If the current row is at the bottom of the triangle ({\tt Row==Tri.length}), then the leaf value is returned. Otherwise, it makes a non-deterministic choice between two branches, one going straight down, and the other going down to the adjacent number. This program is compact, and runs very fast. For the 100-row triangle that is provided by the Euler project,\footnote{http://projecteuler.net/problem=67} this program finds an answer in only 0.01 second on an Intel i5 machine.

The above program can be generalized for classic planning. Given an initial state, a set of final states, and a set of possible actions, the classic planning problem is to find a plan that transforms the initial state to a final state. Figure \ref{fig:plan} shows the framework of a tabled planner.

\begin{figure}[tb]
\begin{center}
\begin{verbatim}
        table (+,-,min)
        plan(S,Plan,Cost),final(S) => Plan=[],Cost=0.
        plan(S,Plan,Cost) =>
            action(S,S1,Action,ActionCost),
            plan(S1,Plan1,Cost1),
            Plan = [Action|Plan1],
            Cost = Cost1+ActionCost.
\end{verbatim}
\end{center}
\caption{\label{fig:plan}The framework of a tabled planner.}
\end{figure}

The call {\tt plan(S,Plan,Cost)} binds {\tt Plan} to an optimal plan that can transform state {\tt S} to a final state. The predicate {\tt final(S)} succeeds if {\tt S} is a final state, and the predicate {\tt action} encodes the set of actions in the problem. The tabled program performs a state-space graph search: for a state that occurs in multiple branches in the search tree, the tabled program only expands it once. This framework demonstrated a surprisingly good performance on the Sokoban problem \cite{ZhouD13}, which was a benchmark used in the ASP and IPC competitions. The same framework was also used in a program for the Petrobras logistic problem \cite{BartakZ14}.

The above framework performs depth-unbounded search. For many planning problems, branch and bound is useful for finding optimal solutions. Another argument can be added to the {\tt plan} predicate in order to indicate the current resource limit. If the resource limit is negative, then the current branch can be pruned. The problem is determining which mode to use for the resource-limit argument. If it is treated as an input argument with the mode (+), then calls with the same state and different resource limits are no longer variants, and will be resolved separately. If the resource limit is passed as an {\tt nt} argument, then the framework no longer guarantees the completeness or soundness, because the {\tt nt} argument is disregarded in variant checking, and once a call is completed with a failure it will fail forever, no matter how big the resource limit is. This problem is nicely fixed by the {\it resource-bounded tabled search} technique, which will be described in the next section.

\section{The {\tt planner} Module of Picat}
Planning has been a target problem for logic programming since its inception. The first logic programming language, PLANNER \cite{Hewitt69}, was designed as ``a language for proving theorems and manipulating models in a robot'', and planning has been an important problem domain for Prolog \cite{Kowalski79,warplan}. Nevertheless, Prolog is not recognized as an effective tool for planning. Answer Set Programming (ASP), which is based on the satisfiability approach to planning \cite{KautzS92,Rintanen12}, has had more successes than Prolog in solving planning problems \cite{Lifschitz02,gekakasc12a}. Other logic-based languages, including action languages \cite{DFP11} and transaction logic \cite{FodorK10}, have also been designed for planning.

The {\tt planner} module of Picat is based on the framework given in Figure \ref{fig:plan}. For a planning problem, users only need to specify conditions on the final states and the set of actions, and call one of the search predicates in the module on an initial state in order to find a plan or an optimal plan. The module provides predicates for both {\it resource-unbounded} search and {\it resource-bounded} search. The following two predicates perform resource-bounded search:

\begin{itemize}
\item \texttt{plan($S$,$Limit$,$Plan$,$PlanCost$)}: This predicate, if it succeeds, binds $Plan$ to a plan that can transform state $S$ to a final state. $PlanCost$ is the cost of $Plan$, which cannot exceed $Limit$, a given non-negative integer. 

\item \texttt{best\_plan($S$,$Limit$,$Plan$,$PlanCost$)}: This predicate iteratively uses {\tt plan/4} to search for an optimal plan, starting with the resource limit 0 and incrementally increasing the limit until a plan is found, or until the resource limit exceeds $Limit$, in which case the call fails. 
\end{itemize}

The implementation of {\tt plan/4} follows the framework in Figure \ref{fig:plan}. The resource limit argument is treated in such a way that it is tabled but not used in variant checking. This predicate searches for a plan by performing {\it resource-bounded} search, which only expands a state if the state is new and its resource limit is non-negative, or if the state has previously failed but the current occurrence has a higher resource limit than before. The implementation of {\tt best\_plan} also takes advantage of the tabled states and their tabled resource limits. Unlike the IDA* search algorithm \cite{Korf85}, which starts a new round from scratch, Picat reuses the states that were tabled in the previous rounds: when the current state does not have a higher resource limit than the most recent occurrence, Picat immediately fails the state.

The {\tt planner} module also provides a function, named {\tt current\_resource()}, which returns the resource limit of the current call to {\tt plan/4}. This amount can be used to check against a heuristic value. If the heuristic estimate of the cost to travel from the current state to a final state is greater than the resource limit, then the current state should be failed.

Figure \ref{fig:ricochet} gives a program for the Ricochet Robots problem \cite{ButoLR05}. Given an $N\times N$ grid board with predefined horizontal and vertical barriers between some of the adjacent board positions, a set of robots of distinct colors on different board positions, and a target position, the goal of the game is to guide a robot of a given color to the target position via a sequence of robot moves. A robot can move horizontally or vertically from its current position. Once a direction is chosen, the robot moves in that direction until encountering an obstacle, i.e. a barrier, another robot, or the edge of the board. This problem is one of the benchmarks used in the ASP'13 Competition \cite{ASP13}. The ASP encoding for the Potassco solver is given in \cite{Gebser13}.

A state is represented by a structure of the following format: 
\begin{tabbing}
aa \= aaa \= aaa \= aaa \= aaa \= aaa \= aaa \kill
\> \> {\tt s([CurLoc|TargetLoc],ORobotLocs)}
\end{tabbing}
where the first argument is a cons that holds the current position and the target position of the target robot, and the second argument is a sorted list of positions of the other robots. A state is final if the current position and the target position are the same. 

Note that colors of robots are not included in the representation, which makes non-target robots indistinguishable during search. This representation facilitates sharing, because lists are sorted and their common suffices are only tabled once. This representation also breaks symmetries. Two configurations of non-target robots are treated as the same if they only differ by robots' colors. This kind of symmetry is not easy to remove when only flat facts are used, as in ASP and PDDL.

The actions are specified with two rules. The first rule chooses a stopping position for the target robot, and moves the target robot there. The predicate {\tt choose\_move\_dest} non-deterministically chooses one of the four directions, and returns the position right before the first obstacle in that direction. On backtracking, it chooses an alternative direction. The second rule chooses a non-target robot to move. 

\begin{figure}[tb]
\begin{center}
\begin{verbatim}
import planner.

main =>
    init_state(S0),
    best_plan(S0,Plan),
    writeln(Plan).

final(s([Loc|Loc],_)) => true.

action(s([From|To],ORobotLocs),NextState,Action,ActionCost) ?=> 
    NextState = $s([Stop|To],ORobotLocs),
    Action = [From|Stop],
    ActionCost = 1,
    choose_move_dest(From,ORobotLocs,Stop).
action(s(FromTo@[From|_],ORobotLocs),NextState,Action,ActionCost) => 
    NextState = $s(FromTo,ORobotLocs2),
    Action = [RFrom|RTo],
    ActionCost = 1,
    select(RFrom, ORobotLocs,ORobotLocs1),
    choose_move_dest(RFrom,[From|ORobotLocs1],RTo),
    ORobotLocs2 = insert_ordered(ORobotLocs1,RTo).
\end{verbatim}
\end{center}
\caption{\label{fig:ricochet}A Picat program for the Ricochet Robots problem.}
\end{figure}

The program can be improved by using a heuristic function. At the end of each rule for {\tt action}, the following condition can be added:
\begin{verbatim}
    current_resource() > heuristic_val(NextState)
\end{verbatim}
This ensures that the resource limit is greater than the estimated number of steps required to transform {\tt NextState} to a final state. For example, the current state is at least three steps away from the final state if the target robot is not in the same row or the same column, and the target position has no obstacle around it.

Picat has demonstrated a surprising performance on many benchmarks. For the four planning benchmarks used in the ASP'13 competition ({\it Nomystery}, {\it Ricochet}, {\it Sokoban}, and {\it Solitaire}), Picat is one to three orders of magnitude faster than Potassco, the winner of the competition. FastDownward, a winner of IPC'11, also competed in the ASP'11 Model\&Solve competition. The competition results on the planning benchmarks showed that FastDownward was not as competitive as the best-performing ASP solvers. On the Ricochet benchmark, both Picat and Potassco solved all 30 instances that were used in the ASP competition; on average, Potassco took 49.5 seconds per instance, while Picat took 9.3 seconds when no heuristic was used, and 2.2 seconds when the above heuristic was used. 

\section{Conclusion}
This paper has presented the Picat language, focusing on its modeling and solving power for combinatorial problems. Lorenz Schiffmann wrote the following in his review of an alpha release of Picat in June 2013, which nicely summarizes the features of Picat: {\it The Picat language is really cool; it's a very usable mix of logic, functional, constraint, and imperative programming. Scripts can be made quite short but also easily readable. And the built-in tabling is really cool for speeding up recursive programs. I think Picat is like a perfect Swiss army knife that you can do anything with.} 

Future work includes engineering an optimizing SAT compiler; applying tabled planning to more domains, including model-checking domains; automatic translation of action languages, such as PDDL and HTN, to Picat; and program analyzers in Picat, both for Picat itself, and for other languages.

\section*{Acknowledgements}
As acknowledged in the User's Guide \cite{PicatGuide}, many people have contributed in one form or another to the Picat project. The author thanks Jonathan Fruhman, Hakan Kjellerstrand, and Yanhong Annie Liu for reviewing this paper. This work was supported in part by the NSF under grant number CCF1018006.


\end{document}